\RequirePackage{ifpdf}
\ifpdf 
\documentclass[pdftex]{sigma}
\else
\documentclass{sigma}
\fi

\begin{document}

\allowdisplaybreaks

\renewcommand{\thefootnote}{$\star$}

\renewcommand{\PaperNumber}{031}

\FirstPageHeading

\ShortArticleName{On Classical Dynamics of Af\/f\/inely-Rigid Bodies Subject to the Kirchhof\/f--Love Constraints}

\ArticleName{On Classical Dynamics of Af\/f\/inely-Rigid Bodies\\
Subject to the Kirchhof\/f--Love Constraints\footnote{This
paper is a contribution to the Proceedings of the Eighth
International Conference ``Symmetry in Nonlinear Mathematical
Physics'' (June 21--27, 2009, Kyiv, Ukraine). The full collection
is available at
\href{http://www.emis.de/journals/SIGMA/symmetry2009.html}{http://www.emis.de/journals/SIGMA/symmetry2009.html}}}

\Author{Vasyl KOVALCHUK}

\AuthorNameForHeading{V. Kovalchuk}

\Address{Institute of Fundamental Technological Research, Polish Academy of Sciences,\\
5$^{B}$~Pawi\'{n}skiego Str., 02-106 Warsaw, Poland}
\Email{\href{mailto:vkoval@ippt.gov.pl}{vkoval@ippt.gov.pl}}

\ArticleDates{Received November 13, 2009, in f\/inal form March 31, 2010;  Published online April 08, 2010}

\Abstract{In this article we consider the af\/f\/inely-rigid body moving in the three-dimen\-sional physical space and subject to the Kirchhof\/f--Love constraints, i.e., while it deforms homogeneously in the two-dimensional central plane of the body it simultaneously performs one-dimensional oscillations orthogonal to this central plane. For the polar decomposition we obtain the stationary ellipsoids as special solutions of the general, strongly nonlinear equations of motion. It is also shown that these solutions are conceptually dif\/ferent from those obtained earlier for the two-polar (singular value) decomposition.}

\Keywords{af\/f\/inely-rigid bodies with degenerate dimension; Kirchhof\/f--Love constraints; polar decomposition; Green deformation tensor; deformation invariants; stationary ellipsoids as special solutions}

\Classification{37N15; 70E15; 70H33; 74A99}

\section{Introduction}

The special interest in the present work is devoted to the classical description of an af\/f\/inely-rigid (homogeneously deforming) mechanical system subject to the Kirchhof\/f--Love constraints. We know that the standard continuum theory as well as some fundamental theories deal with such objects as membranes, plates, discs, etc. So, the main contribution of this work is to present a~toy model for the analytical description of the above-mentioned objects.

The structure of this article is as follows: Firstly, we will present the main notions about the concept of the af\/f\/inely-rigid body, as a generalization of the metrically-rigid one, and of its special case, i.e., the af\/f\/inely-rigid body with degenerate dimension. Secondly, for convenience of the Reader the main results obtained earlier for the case of two-polar (singular value) decomposition are remembered. Thirdly, an alternative (polar) decomposition is introduced and the equations of motion for our toy model are obtained for the general form of the inertial tensor, i.e., when $J_{1}\neq J_{2}\neq J_{3}$. And f\/inally, three main branches of special solutions (stationary ellipsoids) for our strongly nonlinear equations of motion are gathered in the form of Proposition~\ref{proposition1}. Additionally some remarks about the complementarity of the obtained results to those described in our previous work \cite{KovRoz_09} are presented in the Summary.

So, let us remind some basic facts generally concerning the notion of af\/f\/inely-rigid bodies~\cite{JJS_82,JJS_book,all_04,all_05}.

Let $(M,V,\rightarrow)$ be an af\/f\/ine space and $(M,V,\rightarrow,g)$ be the corresponding Euclidean one, where $M$ is a physical space in which the classical system of material points (discrete or continuous) is placed, $V$ is a linear space of translations (free vectors) in $M$, and $g\in V^{\ast}\otimes V^{\ast}$ is the metric tensor. Also let us introduce an af\/f\/ine $(N,U,\rightarrow)$ and the corresponding Euclidean $(N,U,\rightarrow,\eta)$ spaces, where $N$ is the material space of labels which are assigned to every material point of our body in some way, $U$ is the corresponding linear space of translations in $N$, and $\eta\in U^{\ast}\otimes U^{\ast}$ is the metric tensor. Then the position of the $a$-th material point at the time instant $t$ will be denoted by $x(t,a)$ ($x\in M,\ a\in N$) and an af\/f\/ine mapping from the material space into the physical one is as follows:
\begin{gather*}
x^{i}(t,a)=r^{i}(t)+\varphi^{i}{}_{A}(t)a^{A},
\end{gather*}
where $\varphi(t)$ is a linear part of the af\/f\/ine mapping ($\varphi$ is non-singular for any time instant $t$), i.e., $\varphi(t)\in {\rm LI}(U,V)$, where ${\rm LI}(U,V)$ is a manifold of linear isomorphisms from the linear space $U$ into the linear space $V$, $r(t)$ is the radius-vector of the centre of mass of our body if in the material space the position of the centre of mass is $a^{A}=0$. If the system is continuous, then the label $a$ becomes the Lagrangian radius-vector (material variables) and $x$ becomes the Eulerian radius-vector (physical variables). Thus, at any f\/ixed $t\in \mathbb{R}$ the conf\/iguration space $Q$ of our problem is given by the following expression:
\begin{gather*}
Q={\rm AfI}(N,M)=Q_{\rm tr}\times Q_{\rm int}=M\times
{\rm LI}(U,V),
\end{gather*}
where ``tr'' and ``int'' refer to the translational (spatial translations) and internal (rotations and homogeneous deformations) motions respectively.

The considered system is called an af\/f\/inely-rigid body \cite{JJS_82,JJS_book,JJS-VK_03,all_04,all_05}, i.e., during any admissible motion all af\/f\/ine relations between constituents of the body are invariant (the material straight lines remain straight lines, their parallelism is conserved, and all mutual ratios of segments placed on the same straight lines are constant). The conception of the af\/f\/inely-rigid body is a generalization of the usual metrically-rigid body, in which during any admissible motion all distances (metric relations) between constituents of the body are constant (see, for example,~\cite{Arn_78,Gold_50}).

In this article we concentrate mainly on the case of such an af\/f\/inely-rigid body that is subject to the additional constraints, i.e., it can deform homogeneously in the two-dimensional central plane of the body and simultaneously performs one-dimensional oscillations orthogonal to this central plane. Then the material space $N$ is presented as the Cartesian product $\mathbb{R}^{+} \times \mathbb{R}^{2}$ and the group of material transformations has the form $\mathbb{R}^{+} \times {\rm GL}\left(2,\mathbb{R}\right)$, where $\mathbb{R}^{+}$ is the dilatation group in the third dimension and the material transformations in $\mathbb{R}^{2}$ act as in the case of the usual af\/f\/inely-rigid body with degenerate dimension \cite{Roz_05,Roz_PhD}.

We can identify conf\/igurations $\Phi:\mathbb{R}^{3} \rightarrow \mathbb{R}^{3}$ with the pairs $(\varrho,\varphi)$, where $\varphi$ describes the immersion of the central plane in the physical space, i.e., analytically $\varphi^{i}{}_{A}$ is the $3 \times 2$ matrix. An element $(k,B)$ acts on $(\varrho,\varphi)$ as follows:
\begin{gather*}
(k,B)\in\mathbb{R}^{+}\times{\rm GL}(2,\mathbb{R})\colon \quad (\varrho, \varphi) \mapsto (k\varrho, \varphi B).
\end{gather*}
The conservation of orthogonality of the direction of dilatations to the central plane means that the matrix
\begin{gather*}
\Phi = \left[
\begin{array}{ccc}
  \Phi^{1}{}_{1} & \Phi^{1}{}_{2} & \Phi^{1}{}_{3} \\
  \Phi^{2}{}_{1} & \Phi^{2}{}_{2} & \Phi^{2}{}_{3} \\
  \Phi^{3}{}_{1} & \Phi^{3}{}_{2} & \Phi^{3}{}_{3}
\end{array}
\right]
\end{gather*}
fulf\/ils the condition that third column has to be proportional to the vector product of f\/irst and second ones. If we consider $\Phi^{a}{}_{1}$, $\Phi^{b}{}_{2}$, $a,b=1,2,3$, as independent and arbitrary, then
$\Phi^{a}{}_{3}= \ell \ \varepsilon^{a}{}_{bc} \Phi^{b}{}_{1}
\Phi^{c}{}_{2}$, where $\varepsilon_{abc}$ is the completely antisymmetrical Levi-Civita (permutation) symbol, $\ell$ is the parameter which depends both on the variable describing one-dimensional oscillations orthogonal to the central plane of the body and on the ones describing the state of deformation in this central plane, e.g., for the two-polar (singular value) (\ref{eq_6}) and the polar (\ref{eq_15}) decompositions we have respectively that
\begin{gather*}
\ell_{\rm two-polar}=\frac{\varrho}{\lambda\mu},\qquad \ell_{\rm polar}=\frac{\varrho}{\xi\zeta-\alpha^{2}},
\end{gather*}
where the meaning of variables $\lambda$, $\mu$, $\alpha$, $\xi$, $\zeta$, $\varrho$ is clear from the expressions (\ref{eq_7}) and (\ref{eq_17}) below.

The above-described orthogonality is well known in the theory of plates and shells as the Kirchhof\/f--Love condition \cite{Love_96}.

\section{Two-polar decomposition}

In \cite{KovRoz_09} we discussed the language of the two-polar (singular value) decomposition:
\begin{gather}\label{eq_6}
\Phi\left(\overline{k};\lambda,\mu,\varrho;\theta\right)= R\left(\overline{k}\right)D\left(\lambda,\mu,\varrho\right)U\left(\theta\right)^{-1},\qquad
\lambda,\mu,\varrho>0,
\end{gather}
where $R,U\in{\rm SO}(3,\mathbb{R})$ are proper orthogonal matrices (whereas $\overline{k}$ is a rotation vector, i.e., a non-normalized vector codirectional with the rotation axis whose magnitude is equal to the rotation angle) and $D$ is diagonal, i.e.,
\begin{gather}\label{eq_7}
D(\lambda,\mu,\varrho)=\left[
\begin{array}{ccc}
  \lambda & 0 & 0 \\
  0 & \mu & 0 \\
  0 & 0 & \varrho
\end{array}
\right],\qquad U(\theta)^{-1}=\left[
\begin{array}{ccc}
  \cos\theta & \sin\theta & 0 \\
  -\sin\theta & \cos\theta & 0 \\
  0 & 0 & 1
\end{array}
\right].
\end{gather}
Then the co-moving angular velocities for $R$- and $U$-tops \cite{all_04,all_05} are as follows:
\begin{gather*}
\omega= R^{-1}\dot{R}=R^{T}\dot{R}=
\left[
\begin{array}{ccc}
0 & \omega _{3} & -\omega _{2} \\
-\omega _{3} & 0 & \omega _{1} \\
\omega _{2} & -\omega _{1} & 0
\end{array}
\right],\qquad {\omega}^{T}=-\omega,
\end{gather*}
and
\begin{gather*}
\vartheta=U^{-1}\dot{U}=U^{T}\dot{U}=
\dot{\theta}\left[
\begin{array}{ccc}
  0 & -1 & 0 \\
  1 & 0 & 0 \\
  0 & 0 & 0
\end{array}
\right], \qquad
\vartheta^{T}=-\vartheta.
\end{gather*}
For $\dot{\Phi}$ and $\dot{\Phi}^{T}$ we have the following expressions:
\begin{gather*}
\dot{\Phi}=R\big(\dot{D}+\omega D-D\vartheta\big)U^{-1},\qquad
\dot{\Phi}^{T}=U\big(\dot{D}+\vartheta D-D\omega\big)R^{T}.
\end{gather*}

The kinetic energy is assumed to have the usual form (we have only to substitute the constraints):
\begin{gather*}
T=\tfrac{1}{2}{\rm Tr}\big(J\dot{\Phi}^{T}\dot{\Phi}\big)=
\tfrac{1}{2}{\rm Tr}\big(U^{-1}JU\big[\dot{D}+\vartheta D-D\omega\big]\big[\dot{D}+\omega D-D\vartheta\big]\big),
\end{gather*}
where $J\in U\otimes U$ is the twice contravariant, symmetric, non-singular, positively-def\/inite tensor describing the inertial properties of our af\/f\/inely-rigid body. If we take $J$ in the diagonal form $J={\rm Diag}\left(J_{1},J_{2},J_{3}\right)$, then the above kinetic energy can be rewritten as follows:
\begin{gather}
T=\frac{J_{1}\cos^{2}\theta+J_{2}\sin^{2}\theta}{2}\left(\frac{d\lambda}{dt}\right)^{2}+
\frac{J_{1}\sin^{2}\theta+J_{2}\cos^{2}\theta}{2}\left(\frac{d\mu}{dt}\right)^{2}
\nonumber \\
\phantom{T=}{}+ \frac{J_{3}}{2}\left(\frac{d\varrho}{dt}\right)^{2}
+\frac{\left(J_{1}\sin^{2}\theta+J_{2}\cos^{2}\theta\right)\mu^{2}+J_{3}\varrho^{2}}{2}\
\omega_{1}^{2}\nonumber\\
\phantom{T=}{}+
\frac{\left(J_{1}\cos^{2}\theta+J_{2}\sin^{2}\theta\right)\lambda^{2}+J_{3}\varrho^{2}}{2}\ \omega_{2}^{2}
+\left(J_{1}+J_{2}\right)\lambda\mu\omega_{3}\frac{d\theta}{dt}\nonumber\\
\phantom{T=}{}+
\left(J_{1}-J_{2}\right)\sin 2\theta
\left[\left(\mu\frac{d\mu}{dt}-\lambda\frac{d\lambda}{dt}\right)\frac{d\theta}{dt}+
\left(\lambda\frac{d\mu}{dt}-\mu\frac{d\lambda}{dt}\right)\omega_{3}+
\lambda\mu\omega_{1}\omega_{2}\right]\nonumber\\
\phantom{T=}{}+
\frac{\left(J_{1}\cos^{2}\theta+J_{2}\sin^{2}\theta\right)\lambda^{2}
+\left(J_{1}\sin^{2}\theta+J_{2}\cos^{2}\theta\right)\mu^{2}}{2}\ \omega_{3}^{2}\nonumber\\
\phantom{T=}{}+
\frac{\left(J_{1}\sin^{2}\theta+J_{2}\cos^{2}\theta\right)\lambda^{2}+
\left(J_{1}\cos^{2}\theta+J_{2}\sin^{2}\theta\right)\mu^{2}}{2}
\left(\frac{d\theta}{dt}\right)^{2}.\label{eq_12}
\end{gather}
The above expressions signif\/icantly simplify when we consider the isotropic case in the central plane of the body, i.e., when we have $J_{1}=J_{2}=J$. Then
\begin{gather*}
T =  \frac{J}{2}\left[\left(\frac{d\lambda}{dt}\right)^{2}+
\left(\frac{d\mu}{dt}\right)^{2}\right]+
\frac{J_{3}}{2}\left(\frac{d\varrho}{dt}\right)^{2}+
\frac{J\mu^{2}+J_{3}\varrho^{2}}{2}\omega_{1}^{2}\nonumber\\
\phantom{T=}{}+
\frac{J\lambda^{2}+J_{3}\varrho^{2}}{2}\omega_{2}^{2}+2J\lambda \mu \omega_{3}\frac{d\theta}{dt}+\frac{J}{2}\left(\lambda^{2}+\mu^{2}\right)
\left[\omega_{3}^{2}+\left(\frac{d\theta}{dt}\right)^{2}\right].
\end{gather*}
We also remind here that the corresponding expression for the kinetic energy in the canonical variables has the following form:
\begin{gather*}
\mathcal{T} = \frac{s_{1}^{2}}{2\left(J\mu^{2}+J_{3}\varrho^{2}\right)}
+\frac{s_{2}^{2}}{2\left(J\lambda^{2}+J_{3}\varrho^{2}\right)}\nonumber\\
\phantom{\mathcal{T} =}{}
+\frac{\left(\lambda^{2}+\mu^{2}\right)\left(s_{3}^{2}+p_{\theta}^{2}\right)
-4\lambda \mu p_{\theta}s_{3}}{2J\left(\lambda^{2}-\mu^{2}\right)^{2}}
+\frac{p_{\lambda}^{2}+p_{\mu}^{2}}{2J}+\frac{p_{\varrho}^{2}}{2J_{3}}.
\end{gather*}

Then introducing some modelled potentials in \cite{KovRoz_09} we obtained the Hamiltonian (total energy) and calculated the corresponding equations of motion for the isotropic case with the help of the Poisson brackets. In the present article we concentrate mainly on the alternative decomposition, i.e., the polar one. The main advantages of this decomposition are the more physically intuitive division on three main terms in the kinetic energy expression (see the formulas (\ref{eq_19})--(\ref{eq_22}) below) and the possibility to obtain the equations of motion in the quite simple form (see the expressions (\ref{eq_74})--(\ref{eq_80}) below) even for the general case, when the inertial tensor is not isotropic in the central plane ($J_{1}\neq J_{2}$).

\section{Polar decomposition}

Instead of (\ref{eq_6}) we can also use the language of the polar decomposition, i.e.,
\begin{gather}\label{eq_15}
\Phi\left(\overline{\kappa};\alpha,\xi,\zeta,\varrho\right)=
L\left(\overline{\kappa}\right)S\left(\alpha,\xi,\zeta,\varrho\right),
\end{gather}
where $L\in{\rm SO}(3,\mathbb{R})$ is a proper orthogonal matrix and $S\in{\rm Sym(3,\mathbb{R})}$ is symmetrical. The connection between the polar and two-polar decompositions is given by the following expressions:
\begin{gather}
L = RU^{-1},\nonumber\\
\nu = L^{-1}\dot{L}=-\nu^{T}=
\left[
\begin{array}{ccc}
0 & \nu_{3} & -\nu_{2}\\
-\nu_{3} & 0 & \nu_{1}\\
\nu_{2} & -\nu_{1}& 0
\end{array}
\right]=U\left(\omega-\vartheta\right)U^{-1} \nonumber\\
\phantom{\nu}{} = \left[
\begin{array}{ccc}
0 & \omega_{3}+\dot{\theta} &
-\omega_{1}\sin\theta-\omega_{2}\cos\theta\\
-\omega_{3}-\dot{\theta} & 0 &
\omega_{1}\cos\theta-\omega_{2}\sin\theta\\
\omega_{1}\sin\theta+\omega_{2}\cos\theta &
\omega_{2}\sin\theta-\omega_{1}\cos\theta& 0
\end{array}
\right],  \nonumber\\ 
S =
\left[
\begin{array}{ccc}
\xi & \alpha & 0\\
\alpha & \zeta & 0\\
0 & 0& \varrho
\end{array}
\right]=UDU^{-1}
 =
\left[
\begin{array}{ccc}
\lambda\cos^{2}\theta+\mu\sin^{2}\theta &
\left(\lambda-\mu\right)\sin\theta\cos\theta & 0\\
\left(\lambda-\mu\right)\sin\theta\cos\theta & \lambda\sin^{2}\theta+\mu\cos^{2}\theta & 0\\
0 & 0& \varrho
\end{array}
\right],\label{eq_17}
\end{gather}
and then the Green deformation tensor, which is not sensitive with respect to the left orthogonal mappings, is as follows:
\begin{gather*}
G = \Phi^{T}\Phi=S^{2}=
\left[
\begin{array}{ccc}
  \xi^{2}+\alpha^{2} & \left(\xi+\zeta\right)\alpha & 0 \\
  \left(\xi+\zeta\right)\alpha & \zeta^{2}+\alpha^{2} & 0 \\
  0 & 0 & \varrho^{2}
\end{array}
\right]=UD^{2}U^{-1}\nonumber\\
\phantom{G}{} = \left[
\begin{array}{ccc}
  \lambda^{2}\cos^{2}\theta+\mu^{2}\sin^{2}\theta & \left(\lambda^{2}-\mu^{2}\right)\sin\theta\cos\theta & 0 \\
  \left(\lambda^{2}-\mu^{2}\right)\sin\theta\cos\theta & \lambda^{2}\sin^{2}\theta+\mu^{2}\cos^{2}\theta & 0 \\
  0 & 0 & \varrho^{2}
\end{array}
\right],
\end{gather*}
where for the positive def\/initeness the parameters have to fulf\/il the conditions
\begin{alignat*}{3}
&\xi=\lambda\cos^{2}\theta+\mu\sin^{2}\theta>0,\qquad &&
\zeta=\lambda\sin^{2}\theta+\mu\cos^{2}\theta>0, & \\ 
& \xi\zeta-\alpha^{2}=\lambda\mu>0,\qquad && \varrho>0.& 
\end{alignat*}
For the polar decomposition we can as well introduce the concept of deformation invariants $\mathcal{K}^{a}$, $a=1,2,3$, which may be chosen, e.g., as the eigenvalues of the symmetric matrix $G$:
\begin{gather*}
\det\left[G-\mathcal{K}\mathbb{I}_{3}\right]=0,
\end{gather*}
where $\mathbb{I}_{3}$ is the $3\times 3$ identity matrix, and the solutions are as follows:
\begin{gather*}
\mathcal{K}_{1,2}=\tfrac{1}{2}\Big(
\xi^{2}+\zeta^{2}+2\alpha^{2}\pm
\left(\xi+\zeta\right)\sqrt{\left(\xi-\zeta\right)^{2}+4\alpha^{2}}\Big),\qquad
\mathcal{K}_{3}=\varrho^{2}.
\end{gather*}
The above deformation invariants are not sensitive with respect to both the spatial and material rigid rotations (isometries).

Let us consider the Lagrangian $L=T-V\left(\Phi\right)$ and then the Hamiltonian $H=\mathcal{T}+V\left(\Phi\right)$, where the kinetic energy (\ref{eq_12}) can be rewritten for the polar decomposition as follows:
\begin{gather}\label{eq_19}
T=T_{\rm rot}+T_{\rm rot-def}+T_{\rm def},
\end{gather}
where
\begin{gather}
T_{\rm rot} = \frac{J_{1}\alpha^{2}+J_{2}\zeta^{2}+
J_{3}\varrho^{2}}{2}\nu_{1}^{2}+
\frac{J_{1}\xi^{2}+J_{2}\alpha^{2}+
J_{3}\varrho^{2}}{2}\nu_{2}^{2}\nonumber\\
\phantom{T_{\rm rot} =}{}  + \frac{J_{1}\xi^{2}+J_{2}\zeta ^{2}+
\left( J_{1}+J_{2}\right)\alpha^{2}}{2}
\nu_{3}^{2}-\left(J_{1}\xi+J_{2}\zeta\right)
\alpha\nu_{1}\nu_{2}\label{eq_20}
\end{gather}
describes the coupling between the angular velocity $\nu$ of the $L$-top and deformation matrix $S$,
\begin{gather}\label{eq_21}
T_{\rm rot-def}=\left(J_{1}\alpha\frac{d\xi}{dt}-
J_{2}\alpha\frac{d\zeta}{dt}-
\left(J_{1}\xi-J_{2}\zeta\right)\frac{d\alpha}{dt}
\right)\nu_{3}
\end{gather}
describes the connection between the angular and deformation velocities, and f\/inally
\begin{gather}\label{eq_22}
T_{\rm def}=
\frac{J_{1}+J_{2}}{2}\left(\frac{d\alpha}{dt}\right)^{2}+
\frac{J_{1}}{2}\left(\frac{d\xi}{dt}\right)^{2}+
\frac{J_{2}}{2}\left(\frac{d\zeta}{dt}\right)^{2}+
\frac{J_{3}}{2}\left(\frac{d\varrho}{dt}\right)^{2}
\end{gather}
describes the kinetic energy of the deformation oscillations, whereas the potential term $V\left(\Phi\right)$ depends on $\Phi$ only through the Green deformation tensor $G=S^{2}$, i.e., the potential term adapted to the polar decomposition is a function only of $\alpha$, $\xi$, $\zeta$, and $\varrho$.

Performing the Legendre transformation we obtain that
\begin{gather*}
\pi_{1}=\frac{\partial T}{\partial \nu_{1}}=
\left(J_{1}\alpha^{2}+J_{2}\zeta^{2}+
J_{3}\varrho^{2}\right)\nu_{1}-
\left(J_{1}\xi+J_{2}\zeta\right)\alpha\nu_{2},\\ 
\pi_{2}=\frac{\partial T}{\partial \nu_{2}}=
\left(J_{1}\xi^{2}+J_{2}\alpha^{2}+
J_{3}\varrho^{2}\right)\nu_{2}-
\left(J_{1}\xi+J_{2}\zeta\right)\alpha\nu_{1},\\ 
\pi_{3}=\frac{\partial T}{\partial \nu_{3}}=
\left(J_{1}\xi^{2}+J_{2}\zeta^{2}+
\left(J_{1}+J_{2}\right)\alpha^{2}\right)\nu_{3}
+J_{1}\alpha\dot{\xi}-J_{2}\alpha\dot{\zeta}-
\left(J_{1}\xi-J_{2}\zeta\right)\dot{\alpha},\\ 
p_{\alpha}=\frac{\partial T}{\partial \dot{\alpha}}=\left( J_{1}+J_{2}\right)\dot{\alpha}-
\left(J_{1}\xi-J_{2}\zeta\right)\nu_{3},\\ 
p_{\xi}=\frac{\partial T}{\partial \dot{\xi}}=
J_{1}\left(\dot{\xi}+\alpha\nu_{3}\right),\\ 
p_{\zeta}=\frac{\partial T}{\partial \dot{\zeta}}=J_{2}\left(\dot{\zeta}-\alpha\nu_{3}\right),\\ 
p_{\varrho}=\frac{\partial T}{\partial \dot{\varrho}}=J_{3}\dot{\varrho},
\end{gather*}
where $\pi_{i}$ are canonical ``spin'' variables conjugate to angular velocities $\nu_{i}$.

Therefore after inverting the above dependencies, i.e.,
\begin{gather*}
\nu_{1} = \frac{\left(J_{1}\xi^{2}+J_{2}\alpha^{2}+
J_{3}\varrho^{2}\right)\pi_{1}+
\left(J_{1}\xi+J_{2}\zeta\right)\alpha\pi_{2}}{J_{1}
J_{2}\left(\alpha^{2}-\xi\zeta\right)^{2}+
\left[J_{1}\xi^{2}+J_{2}\zeta^{2}+
\left(J_{1}+J_{2}\right)\alpha^{2}\right]J_{3}\varrho^{2}+
J^{2}_{3}\varrho^{4}},\\ 
\nu_{2} = \frac{\left(J_{1}\xi+
J_{2}\zeta\right)\alpha\pi_{1}+
\left(J_{1}\alpha^{2}+J_{2}\zeta^{2}+
J_{3}\varrho^{2}\right)\pi_{2}}{J_{1}
J_{2}\left(\alpha^{2}-\xi\zeta\right)^{2}+
\left[J_{1}\xi^{2}+J_{2}\zeta^{2}+
\left(J_{1}+J_{2}\right)\alpha^{2}\right]J_{3}\varrho^{2}+
J^{2}_{3}\varrho^{4}},\\ 
\nu_{3} = \frac{\left(J_{1}+J_{2}\right)
\left[\pi_{3}+\alpha\left(p_{\zeta}-p_{\xi}\right)\right]+
\left(J_{1}\xi-J_{2}\zeta\right)p_{\alpha}}{J_{1}J_{2}
\left(\xi+\zeta\right)^{2}},\\ 
\frac{d\alpha}{dt} = \frac{\left(J_{1}\xi-J_{2}\zeta\right)\left[\pi_{3}+
\alpha\left(p_{\zeta}-p_{\xi}\right)\right]+\left(J_{1}\xi^{2}+
J_{2}\zeta^{2}\right)p_{\alpha}}{J_{1}J_{2}
\left(\xi+\zeta\right)^{2}},\\ 
\frac{d\xi}{dt} = \frac{p_{\xi}}{J_{1}}-
\alpha\frac{\left(J_{1}+J_{2}\right)
\left[\pi_{3}+\alpha\left(p_{\zeta}-p_{\xi}\right)\right]+
\left(J_{1}\xi-J_{2}\zeta\right)p_{\alpha}}{J_{1}J_{2}
\left(\xi+\zeta\right)^{2}},\\ 
\frac{d\zeta}{dt} = \frac{p_{\zeta}}{J_{2}}+
\alpha\frac{\left(J_{1}+J_{2}\right)
\left[\pi_{3}+\alpha\left(p_{\zeta}-p_{\xi}\right)\right]+
\left(J_{1}\xi-J_{2}\zeta\right)p_{\alpha}}{J_{1}J_{2}
\left(\xi+\zeta\right)^{2}},\\ 
\frac{d\varrho}{dt} = \frac{p_{\varrho}}{J_{3}},
\end{gather*}
we obtain the kinetic energy in the canonical variables as follows:
\begin{gather}
\mathcal{T} =
\frac{\left(J_{1}\xi^{2}+J_{2}\alpha^{2}+
J_{3}\varrho^{2}\right)\pi^{2}_{1}+
\left(J_{1}\alpha^{2}+J_{2}\zeta^{2}+
J_{3}\varrho^{2}\right)\pi^{2}_{2}
}{2\left(J_{1}
J_{2}\left(\alpha^{2}-\xi\zeta\right)^{2}+
\left[J_{1}\xi^{2}+J_{2}\zeta^{2}+
\left(J_{1}+J_{2}\right)\alpha^{2}\right]J_{3}\varrho^{2}+
J^{2}_{3}\varrho^{4}\right)}
\nonumber\\
 \phantom{\mathcal{T} = }{} + \frac{\left(J_{1}\xi+J_{2}\zeta\right)\alpha\pi_{1}\pi_{2}
}{J_{1}
J_{2}\left(\alpha^{2}-\xi\zeta\right)^{2}+
\left[J_{1}\xi^{2}+J_{2}\zeta^{2}+
\left(J_{1}+J_{2}\right)\alpha^{2}\right]J_{3}\varrho^{2}+
J^{2}_{3}\varrho^{4}}\nonumber\\
\phantom{\mathcal{T} = }{} + \frac{J_{1}+J_{2}}{2J_{1}J_{2}\left(\xi+\zeta\right)^{2}}
\left[\pi_{3}+\alpha\left(p_{\zeta}-p_{\xi}\right)\right]^{2}
+\frac{J_{1}\xi^{2}+J_{2}\zeta^{2}}{2J_{1}J_{2}\left(\xi+\zeta\right)^{2}}
p^{2}_{\alpha}\nonumber\\
\phantom{\mathcal{T} = }{} + \frac{J_{1}\xi-J_{2}\zeta}{J_{1}J_{2}
\left(\xi+\zeta\right)^{2}}\left[\pi_{3}+\alpha\left(p_{\zeta}-
p_{\xi}\right)\right]p_{\alpha}
+\frac{p^{2}_{\xi}}{2J_{1}}
+\frac{p^{2}_{\zeta}}{2J_{2}}
+\frac{p^{2}_{\varrho}}{2J_{3}}.\label{eq_73}
\end{gather}
From the above kinetic energy expressions (\ref{eq_19})--(\ref{eq_22}) one can see that the generalized velocities $\dot{\alpha}$, $\dot{\xi}$, $\dot{\zeta}$ corresponding to $\alpha$, $\xi$, $\zeta$ and other variables describing the motion in the central plane of the body are separated from the generalized velocity $\dot{\varrho}$ describing the one-dimensional oscillations orthogonal to this central plane. The same can be said also about the above expression in the canonical variables (\ref{eq_73}), i.e., the momentum $p_{\varrho}$ conjugated to $\varrho$ is orthogonal (in the sense of metrics encoded in the kinetic energy expression) to the other canonical momenta. Hence, the most simple are those dynamical models in which also the isotropic potential will have the separated form:
\begin{gather*}
V\left(\alpha,\xi,\zeta,\varrho\right)=
V_{\rm plane}\left(\alpha,\xi,\zeta\right)+V_{\varrho}\left(\varrho\right),
\end{gather*}
where as the potential $V_{\varrho}$ we can take, e.g., the following potential which describes the nonlinear oscillations and is in accordance with the main demands of the elasticity theory, i.e.,
\begin{gather*}
V_{\varrho}(\varrho)= \frac{a}{\varrho}+ \frac{b}{2}\varrho^{2},\qquad a,b>0,
\end{gather*}
where the f\/irst term prevents from the unlimited compressing of the body, whereas the second one restricts the motion for large values of $\varrho$, i.e., prevents from the non-physical unlimited stretching of the body.

So, the Hamiltonian (total energy) can be written as follows:
\begin{gather*}
H=\mathcal{T}+V_{\rm plane}\left(\alpha,\xi,\zeta\right)+V_{\varrho}(\varrho),
\end{gather*}
where $\mathcal{T}$ is taken in the form of (\ref{eq_73}). Then the equations of motion can be calculated with the help of the following Poisson brackets:
\begin{gather*}
\frac{d\pi_{i}}{dt}=\left\{\pi_{i},H\right\},\qquad
\frac{dp_{\alpha}}{dt}=\left\{p_{\alpha},H\right\},\qquad
\frac{dp_{\xi}}{dt}=\left\{p_{\xi},H\right\},\nonumber\\
\frac{dp_{\zeta}}{dt}=\left\{p_{\zeta},H\right\},\qquad
\frac{dp_{\varrho}}{dt}=\left\{p_{\varrho},H\right\}. 
\end{gather*}
The only non-zero basic Poisson brackets are
\begin{gather*}
\left\{\alpha,p_{\alpha}\right\}=\left\{\xi,p_{\xi}\right\}=
\left\{\zeta,p_{\zeta}\right\}=\left\{\varrho,p_{\varrho}\right\}=1,\qquad
\left\{\pi_{i},\pi_{j}\right\}=-\varepsilon_{ij}{}^{k}\pi_{k},
\end{gather*}
where the former expressions follow directly from the def\/inition of the Poisson
bracket and the latter ones are based on the structure constants of the special orthogonal group ${\rm SO}\left(3,\mathbb{R}\right)$.

First of all, let us rewrite the kinetic energy (\ref{eq_73})
in a more symbolic way, i.e.,
\begin{gather*}
\mathcal{T}=\frac{\Omega\left(\pi_{1},\pi_{2}\right)}{2\Xi}
+\frac{\Upsilon\left(\pi_{3}+\alpha\left(p_{\zeta}-p_{\xi}\right),
p_{\alpha}\right)}{2J_{1}J_{2}\left(\xi+\zeta\right)^{2}}
+\frac{p^{2}_{\xi}}{2J_{1}}
+\frac{p^{2}_{\zeta}}{2J_{2}}
+\frac{p^{2}_{\varrho}}{2J_{3}},
\end{gather*}
where
\begin{gather*}
\Xi=
J_{1}J_{2}\left(\alpha^{2}-\xi\zeta\right)^{2}+
\left[J_{1}\xi^{2}+J_{2}\zeta^{2}+
\left(J_{1}+J_{2}\right)\alpha^{2}\right]J_{3}\varrho^{2}+
J^{2}_{3}\varrho^{4},
\end{gather*}
and two expressions built of the canonical momenta are as follows:
\begin{gather*}
\Omega\left(\pi_{1},\pi_{2}\right) = \left(J_{1}\xi^{2}+J_{2}\alpha^{2}
+J_{3}\rho^{2}\right)\pi^{2}_{1}+
2\left(J_{1}\xi+J_{2}\zeta\right)\alpha\pi_{1}\pi_{2}\nonumber\\
\phantom{\Omega\left(\pi_{1},\pi_{2}\right) =}{}
+\left(J_{1}\alpha^{2}+J_{2}\zeta^{2}+J_{3}\rho^{2}
\right)\pi^{2}_{2},\\ 
\Upsilon\left(\pi_{3}+\alpha\left(p_{\zeta}-p_{\xi}\right),p_{\alpha}\right)=
\left(J_{1}+J_{2}\right)
\left[\pi_{3}+\alpha\left(p_{\zeta}-p_{\xi}\right)\right]^{2}
+\left(J_{1}\xi^{2}+J_{2}\zeta^{2}\right)p^{2}_{\alpha}\nonumber\\
\phantom{\Upsilon\left(\pi_{3}+\alpha\left(p_{\zeta}-p_{\xi}\right),p_{\alpha}\right)=}{}
+2\left(J_{1}\xi-J_{2}\zeta\right)\left[\pi_{3}+\alpha\left(p_{\zeta}-
p_{\xi}\right)\right]p_{\alpha}.
\end{gather*}
Then we obtain the following equations of motion:
\begin{gather}
\frac{d\pi_{1}}{dt} = -\frac{\left[\left(J_{1}\xi+J_{2}\zeta\right)
\alpha\pi_{1}+\left(J_{1}\alpha^{2}+J_{2}\zeta^{2}+J_{3}\varrho^{2}\right)
\pi_{2}\right]\pi_{3}}{\Xi}\nonumber\\
 \phantom{\frac{d\pi_{1}}{dt} =}{} + \frac{\pi_{2}\left[\left(J_{1}+J_{2}\right)\left[\pi_{3}+\alpha\left(p_{\zeta}
-p_{\xi}\right)\right]+\left(J_{1}\xi-J_{2}\zeta\right)p_{\alpha}
\right]}{J_{1}J_{2}\left(\xi+\zeta\right)^{2}},\label{eq_74}\\
\frac{d\pi_{2}}{dt} = \frac{\left[\left(J_{1}\xi^{2}+J_{2}\alpha^{2}+
J_{3}\varrho^{2}\right)\pi_{1}+\left(J_{1}\xi+J_{2}\zeta\right)
\alpha\pi_{2}\right]\pi_{3}}{\Xi}\nonumber\\
\phantom{\frac{d\pi_{2}}{dt} =}{}  - \frac{\pi_{1}\left[\left(J_{1}+J_{2}\right)\left[\pi_{3}+\alpha\left(p_{\zeta}
-p_{\xi}\right)\right]+\left(J_{1}\xi-J_{2}\zeta\right)p_{\alpha}
\right]}{J_{1}J_{2}\left(\xi+\zeta\right)^{2}},\label{eq_75}\\
\frac{d\pi_{3}}{dt} = \frac{\left(J_{1}\xi+J_{2}\zeta\right)
\alpha\left(\pi^{2}_{1}-\pi^{2}_{2}\right)+\left[J_{1}\left(\alpha^{2}-
\xi^{2}\right)+J_{2}\left(\zeta^{2}-\alpha^{2}\right)\right]
\pi_{1}\pi_{2}}{\Xi},\quad \label{eq_76}\\
\frac{d\alpha}{dt} = -\frac{\partial V_{\rm plane}}{\partial \alpha}
-\frac{\left(J_{2}\pi^{2}_{1}+J_{1}\pi^{2}_{2}\right)\alpha
+\left(J_{1}\xi+J_{2}\zeta\right)\pi_{1}\pi_{2}}{\Xi}\nonumber\\
\phantom{\frac{d\alpha}{dt} =}{}
 + \frac{2J_{1}J_{2}\alpha\left(\alpha^{2}-\xi\zeta\right)+\left(J_{1}+
J_{2}\right)\alpha J_{3}\varrho^{2}}{\Xi^{2}}\Omega\left(\pi_{1},\pi_{2}\right)\nonumber\\
\phantom{\frac{d\alpha}{dt} =}{}
 - \frac{\left(J_{1}+J_{2}\right)\left[\pi_{3}+\alpha\left(
p_{\zeta}-p_{\xi}\right)\right]+\left(J_{1}\xi-J_{2}\zeta\right) p_{\alpha}}{J_{1}J_{2}\left(\xi+\zeta\right)^{2}}
\left(p_{\zeta}-p_{\xi}\right),\label{eq_77}\\
\frac{d\xi}{dt} = -\frac{\partial V_{\rm plane}}{\partial \xi}
-\frac{J_{1}\xi\pi^{2}_{1}+J_{1}\alpha\pi_{1}\pi_{2}}{\Xi}
 + \frac{J_{1}J_{2}\zeta\left(\xi\zeta-\alpha^{2}\right)+J_{1}\xi J_{3}\varrho^{2}}{\Xi^{2}}\Omega\left(\pi_{1},\pi_{2}\right)\nonumber\\
 \phantom{\frac{d\xi}{dt} =}{}
 - \frac{J_{1}\xi p^{2}_{\alpha}+J_{1}\left[\pi_{3}+\alpha\left(
p_{\zeta}-p_{\xi}\right)\right]p_{\alpha}}{J_{1}J_{2}\left(\xi+\zeta\right)^{2}}
+\frac{\Upsilon\left(\pi_{3}+\alpha\left(p_{\zeta}-p_{\xi}\right),p_{\alpha}
\right)}{J_{1}J_{2}\left(\xi+\zeta\right)^{3}},\label{eq_78}\\
\frac{d\zeta}{dt} = -\frac{\partial V_{\rm plane}}{\partial \zeta}
-\frac{J_{2}\zeta\pi^{2}_{2}+J_{2}\alpha\pi_{1}\pi_{2}}{\Xi}
 + \frac{J_{1}\xi J_{2}\left(\xi\zeta-\alpha^{2}\right)+J_{2}\zeta J_{3}\varrho^{2}}{\Xi^{2}}\Omega\left(\pi_{1},\pi_{2}\right)\nonumber\\
\phantom{\frac{d\zeta}{dt} =}{}
 - \frac{J_{2}\zeta p^{2}_{\alpha}-J_{2}\left[\pi_{3}+\alpha\left(
p_{\zeta}-p_{\xi}\right)\right]p_{\alpha}}{J_{1}J_{2}\left(\xi+\zeta\right)^{2}}
+\frac{\Upsilon\left(\pi_{3}+\alpha\left(p_{\zeta}-p_{\xi}\right),p_{\alpha}
\right)}{J_{1}J_{2}\left(\xi+\zeta\right)^{3}},\label{eq_79}\\
\frac{d\varrho}{dt} = -\frac{dV_{\varrho}}{d\varrho}
-\frac{J_{3}\rho}{\Xi}\left(\pi^{2}_{1}+\pi^{2}_{2}\right)
 +
\frac{J_{3}\varrho}{\Xi^{2}}\left[J_{1}
\xi^{2}+J_{2}\zeta^{2}+\left(J_{1}+J_{2}\right)\alpha^{2}
+2J_{3}\rho^{2}\right]\Omega\left(\pi_{1},\pi_{2}\right).
\label{eq_80}
\end{gather}
The structure of the above expressions implies that even in the simplest case of the completely separated potential the dynamical coupling between the parameter describing one-dimensional oscillations orthogonal to the central plane of the body and the variables living in this central plane is present.

\section{Stationary ellipsoids as special solutions}

Our equations of motion (\ref{eq_74})--(\ref{eq_80}) are strongly nonlinear and in a~general case there is hardly a~hope to solve them analytically. Nevertheless, there
exists a~way for imaging some features of the phase portrait of such a dynamical system, i.e., we have to f\/ind some special solutions, namely, the stationary ellipsoids \cite{JJS_82,JJS_book}, which are analogous to the ellipsoidal f\/igures of equilibrium well known in astro- \cite{Bog_85} and geophysics, e.g., in the theory of the Earth's shape \cite{Chan_69}.

In the case of the two-polar (singular value) decomposition (\ref{eq_6}) we obtained the above-mentioned special solutions just putting the deformation invariants $\lambda$, $\mu$, $\varrho$ and the angular velocities $\omega$, $\vartheta$ equal to some constant values \cite{KovRoz_09}. But now, in the case of the polar decomposition (\ref{eq_15}), we see that the Green deformation tensor  $G$, therefore the deformation matrix $S$, and the angular velocity $\nu$ of the $L$-top have to be constant \cite{JJS_book}, i.e.,
\begin{gather*}
\frac{dG}{dt}=\frac{d}{dt}\left(\Phi^{T}\Phi\right)=
\frac{d}{dt}\left(S^{2}\right)=0,\qquad
\frac{d\nu}{dt}=
\frac{d}{dt}\big(L^{-1}\dot{L}\big)=0.
\end{gather*}
This means that the $L$-top performs the stationary rotation, i.e., if at the initial time $t=0$ we have that the conf\/iguration of the body is $L_{0}$, then at the time instant $t$ the conf\/iguration will be as follows:
\begin{gather*}
L_{0}\circ e^{\nu t},
\end{gather*}
where $\circ$ is the function composition symbol. We see that the whole af\/f\/inely-rigid body, which at the initial time $t=0$ has the internal conf\/iguration $\Phi_{0}=L_{0}\circ S$, at the time instant $t$ will be in the following conf\/iguration:
\begin{gather}\label{eq_39}
\Phi(t)=L_{0}\circ e^{\nu t}\circ S=e^{\widehat{\nu} t}\circ L_{0}\circ S=e^{\widehat{\nu} t}\circ \Phi_{0},
\end{gather}
where $\widehat{\nu}=L_{0}\circ \nu\circ L^{-1}_{0}$.

\begin{proposition}\label{proposition1}
While the affinely-rigid body rotates in the stationary way around the axis fixed in the physical and material spaces, the deformation and the angular velocity of rotation are not independent and related by some algebraic expressions.
\end{proposition}

\begin{proof}
The trajectories of type (\ref{eq_39}) are the orbits of the Euler (spatial) action of the one-parameter orthogonal group $\left\{e^{\widehat{\nu} t}:t\in\mathbb{R}\right\}\subset {\rm SO}\left(V,g\right)$. Nevertheless, during the motion the conf\/igurations of the body are deformed. At the same time the Green deformation tensor does not perform any oscillations, it is constant. This means that an equilibrium is set between the centrifugal forces coming from the rotation of the body and the elastic forces coming from the fact that $S\neq {\rm Id}_{U}$. This kind of equilibrium is possible only if between the constant values of~$\nu$, $S$~are set some algebraic relations that guarantee the balance of the above-described forces. These algebraic relations between~$\nu$,~$S$ obviously come from the equations of motion (\ref{eq_74})--(\ref{eq_80}). Hence, we propose to divide them into the following three main branches:

$(i)$~ $\nu_{1}\neq 0$, $\nu_{2}=\nu_{3}=0$, then $\pi_{1},\pi_{2}\neq 0$ and $\pi_{3}=p_{\alpha}=p_{\xi}=p_{\zeta}=p_{\varrho}=0$;

$(ii)$ $\nu_{2}\neq 0$, $\nu_{1}=\nu_{3}=0$, then $\pi_{1},\pi_{2}\neq 0$ and $\pi_{3}=p_{\alpha}=p_{\xi}=p_{\zeta}=p_{\varrho}=0$.

For the f\/irst two cases the relations take the same form, i.e.,
\begin{gather}
\frac{\partial V_{\rm plane}}{\partial \alpha}=
-\frac{\left(J_{2}\pi^{2}_{1}+J_{1}\pi^{2}_{2}\right)\alpha
+\left(J_{1}\xi+J_{2}\zeta\right)\pi_{1}\pi_{2}}{\Xi}\nonumber\\
\phantom{\frac{\partial V_{\rm plane}}{\partial \alpha}=}{} +\frac{2J_{1}J_{2}\alpha\left(\alpha^{2}-\xi\zeta\right)+\left(J_{1}+
J_{2}\right)\alpha J_{3}\varrho^{2}}{\Xi^{2}}\Omega\left(\pi_{1},\pi_{2}\right),\label{eq_91}\\
\frac{\partial V_{\rm plane}}{\partial \xi}=
-\frac{J_{1}\xi\pi^{2}_{1}+J_{1}\alpha\pi_{1}\pi_{2}}{\Xi}
+\frac{J_{1}J_{2}\zeta\left(\xi\zeta-\alpha^{2}\right)+J_{1}\xi J_{3}\varrho^{2}}{\Xi^{2}}\Omega\left(\pi_{1},\pi_{2}\right),\label{eq_92}\\
\frac{\partial V_{\rm plane}}{\partial \zeta}=
-\frac{J_{2}\zeta\pi^{2}_{2}+J_{2}\alpha\pi_{1}\pi_{2}}{\Xi}
+\frac{J_{1}\xi J_{2}\left(\xi\zeta-\alpha^{2}\right)+J_{2}\zeta J_{3}\varrho^{2}}{\Xi^{2}}\Omega\left(\pi_{1},\pi_{2}\right),\label{eq_93}\\
\frac{dV_{\varrho}}{d\varrho}=
-\frac{J_{3}\rho}{\Xi}\left(\pi^{2}_{1}+\pi^{2}_{2}\right)
+
\frac{J_{3}\varrho}{\Xi^{2}}\left[J_{1}
\xi^{2}+J_{2}\zeta^{2}+\left(J_{1}+J_{2}\right)\alpha^{2}
+2J_{3}\rho^{2}\right]\Omega\left(\pi_{1},\pi_{2}\right),
\label{eq_94}
\end{gather}
with the compatibility condition
\begin{gather*}
\left(J_{1}\xi+J_{2}\zeta\right)\alpha\left(\pi^{2}_{1}-
\pi^{2}_{2}\right)+\left[J_{1}\left(\alpha^{2}-
\xi^{2}\right)+J_{2}\left(\zeta^{2}-\alpha^{2}\right)\right]
\pi_{1}\pi_{2}=0.
\end{gather*}
We see that, while our parameters $\nu_{1}$ or $\nu_{2}$ take completely arbitrary constant values, the above equations (\ref{eq_91})--(\ref{eq_94}) describe their interrelation with the elements of the symmetrical matrix $S$, i.e., with $\alpha$, $\xi$, $\zeta$, $\varrho$.

$(iii)$ $\nu_{3}\neq 0$, $\nu_{1}=\nu_{2}=0$, then $\pi_{3},p_{\alpha},p_{\xi},p_{\zeta}\neq 0$ and $\pi_{1}=\pi_{2}=p_{\varrho}=0$, whereas
\begin{gather*}
p_{\alpha} = \frac{\left(J_{2}\zeta-J_{1}\xi\right)\pi_{3}}{J_{1}\xi^{2}
+J_{2}\zeta^{2}+\left(J_{1}+J_{2}\right)\alpha^{2}},\\ 
p_{\xi}  = \frac{J_{1}\alpha\pi_{3}}{J_{1}\xi^{2}
+J_{2}\zeta^{2}+\left(J_{1}+J_{2}\right)\alpha^{2}},\\ 
p_{\zeta} = -\frac{J_{2}\alpha\pi_{3}}{J_{1}\xi^{2}
+J_{2}\zeta^{2}+\left(J_{1}+J_{2}\right)\alpha^{2}}.
\end{gather*}
So, for the third case we obtain the following relations:
\begin{gather}
\frac{\partial V_{\rm plane}}{\partial \alpha} =
\frac{\left(J_{1}+J_{2}\right)\left[\pi_{3}+\alpha\left(
p_{\zeta}-p_{\xi}\right)\right]+\left(J_{1}\xi-J_{2}\zeta\right) p_{\alpha}}{J_{1}J_{2}\left(\xi+\zeta\right)^{2}}
\left(p_{\xi}-p_{\zeta}\right),\qquad \label{eq_95}\\
\frac{\partial V_{\rm plane}}{\partial \xi} =
-\frac{J_{1}\xi p^{2}_{\alpha}+J_{1}\left[\pi_{3}+\alpha\left(
p_{\zeta}-p_{\xi}\right)\right]p_{\alpha}}{J_{1}J_{2}\left(\xi+\zeta\right)^{2}}
+\frac{\Upsilon\left(\pi_{3}+\alpha\left(p_{\zeta}-p_{\xi}\right),p_{\alpha}
\right)}{J_{1}J_{2}\left(\xi+\zeta\right)^{3}},\label{eq_96}\\
\frac{\partial V_{\rm plane}}{\partial \zeta} =
-\frac{J_{2}\zeta p^{2}_{\alpha}-J_{2}\left[\pi_{3}+\alpha\left(
p_{\zeta}-p_{\xi}\right)\right]p_{\alpha}}{J_{1}J_{2}\left(\xi+\zeta\right)^{2}}
+\frac{\Upsilon\left(\pi_{3}+\alpha\left(p_{\zeta}-p_{\xi}\right),p_{\alpha}
\right)}{J_{1}J_{2}\left(\xi+\zeta\right)^{3}},\label{eq_97}\\
\frac{dV_{\varrho}}{d\varrho} = 0.\label{eq_98}
\end{gather}
This time our parameter $\nu_{3}$ has a completely arbitrary constant value and the above equations (\ref{eq_95})--(\ref{eq_98}) describe the way in which $\alpha$, $\xi$, $\zeta$, $\varrho$ are related to it.
\end{proof}

\begin{remark}
It should be mentioned that the name ``stationary ellipsoids'' is not the most adequate for the description of the above-obtained stationary solutions. Of course, for the non-restricted af\/f\/inely-rigid body we can visualize this kind of special solutions as follows:
\begin{itemize}\itemsep=0pt
\item At the beginning the body stays in the equilibrium conf\/iguration.

\item Then we switch on some mechanical device which deforms our body in the homogeneous way, i.e., this deformation is the superposition of three material stretchings with the coef\/f\/icients $D_{1}$, $D_{2}$, $D_{3}$, where $D_{i}$, $i=1,2,3$, are the diagonal elements of the deformation matrix $D$ in the two-polar decomposition $\Phi=RDU^{-1}$.

\item After this we start to rotate our mechanical device with the constant angular velocity $\vartheta$ around one of the main axes of the Green deformation tensor $G=\Phi^{T}\Phi=UD^{2}U^{-1}$ so that the state of material deformation follows this movement of the device with the same angular velocity (note that our body itself does not rotate!).

\item And f\/inally, the whole system consisting of the already rotating mechanical device and the body starts also to rotate with the constant angular velocity $\omega$ around the corresponding main axis of the Cauchy deformation tensor $C=\Phi^{-1T}\Phi^{-1}=RD^{2}R^{-1}$ (this time both the state of deformation and our body rotate!).
\end{itemize}
We see that the whole system reminds the gimbals equipped with additional mechanical device deforming the body.

If our parameters $D_{1}$, $D_{2}$, $D_{3}$, $\vartheta$, $\omega$ are chosen in such a way that they fulf\/ill the algebraic relations obtained from the equations of motion, then even when we switch of\/f the mechanical device which generates the state of deformation in the material of our body, nothing will change, i.e., the stretchings will continue to rotate with the same constant angular velocity $\vartheta$ around the same axis in the material and the body will be rotating with the same angular velocity $\omega$ around the same axis in the space. The deformation invariants $D_{1}$, $D_{2}$, $D_{3}$ also will be constant during the above-described two types of rotation.

But in our case of the af\/f\/inely-rigid body subject to the Kirchhof\/f--Love constraints we have the homogeneous deformation only in the central plane of the body, whereas in the perpendicular direction the body performs some nonlinear oscillations. Hence, in this situation the more appropriate name for our special solutions is ``elliptical'', but we have kept the generic name ``ellipsoidal'' for the matter of convenience.
\end{remark}

\section{Summary}

It is interesting to note that the special solutions obtained for the polar decomposition case are conceptually dif\/ferent from those obtained for the two-polar one \cite{KovRoz_09} because here the Green deformation tensor $G=S^{2}$ has a constant value (i.e., $\dot{G}=2S\dot{S}=0$) contrary to the situation described in \cite{KovRoz_09} when the Green deformation tensor $G=\Phi^{T}\Phi=UD^{2}U^{-1}$, as well as the Cauchy one $C=\Phi^{-1T}\Phi^{-1}=RD^{2}R^{-1}$, depended on time explicitly through the time dependence of $U$ and $R$ respectively, i.e.,
\begin{gather*}
\frac{dG}{dt}=U\left(\vartheta D^{2}-D^{2}\vartheta\right)U^{-1}\neq 0,\qquad \frac{dC}{dt}=R\left(\omega D^{2}-D^{2}\omega\right)R^{-1}\neq 0,
\end{gather*}
and performed the stationary rotations around their principal axes, whereas the deformation invariants $\lambda$, $\mu$, $\varrho$ had the constant values.

So, if we additionally keep in mind that in \cite{KovRoz_09} we obtained the stationary solutions only for the isotropic model $J_{1}=J_{2}=J$ and here the general situation $J_{1}\neq J_{2}$ is allowed, then we can compare the four (one in \cite{KovRoz_09} and three here) studied cases according to the following scheme:
\begin{itemize}\itemsep=0pt
\item The only degrees of freedom we can manipulate are the rotational degrees of $R$- and $U$-tops, because the deformation matrix $D$ is constant for this type of stationary solutions.

\item To achieve the constant behaviour of the Green deformation tensor $G=S^{2}=\left(UDU^{-1}\right)^{2}=UD^{2}U^{-1}$ we have to suppose that the $U$-top is f\/ixed and does not rotate at all. If $U$ is constant, then the principal axes of the $R$- and $L\left(=RU^{-1}\right)$-tops (for the two-polar and polar decompositions respectively) rotate in the same manner, i.e., at any moment ones can be obtained from others with the help of applying some constant orthogonal transformation. This situation corresponds to the above-mentioned three cases $(i)$--$(iii)$ describing the stationary rotations of the $L$-top around its three principal axes.

\item If $U$-top is not f\/ixed, then the Green deformation is not constant and we have to consider three branches of the stationary motion for $R$- and $U$-tops when they rotate not independently but in the correlated manner, i.e., either both around their f\/irst principal axes or both around the second ones or both around the third ones \cite{JJS_book}. Nevertheless, for our af\/f\/inely-rigid body subject to the Kirchhof\/f--Love constraints only the third case is possible and exactly this situation was studied in the previous paper \cite{KovRoz_09}.
\end{itemize}
Hence, we see that in the above-described sense the results obtained in this article are essentially dif\/ferent from and simultaneously complementary to those obtained in \cite{KovRoz_09}.

Let us mention that the af\/f\/ine models of degrees of freedom for structured bodies have been studied by many authors. The thorough analysis of some stationary motions for af\/f\/ine bodies and their stability was presented in \cite{GrNa_91,LeSi_90,NoRe_98,NoRe_01,Ru_85}. However, in this article we have discussed other problems.

\subsection*{Acknowledgements}

This paper contains results obtained within the framework of the research project 501 018 32/1992 f\/inanced from the Scientif\/ic Research Support Fund in 2007--2010. The author is greatly indebted to the Ministry of Science and Higher Education for this f\/inancial support.

The author is also very grateful to the referees for their valuable remarks and comments concerning this article and some propositions of the further investigation of the subject.

\newpage

\pdfbookmark[1]{References}{ref}
\LastPageEnding


\begin{thebibliography}{99}

\footnotesize\itemsep=0pt

\bibitem{Arn_78}
Arnold V.I.,
Mathematical methods of classical mechanics, {\it Springer Graduate Texts in Mathematics}, Vol.~60, Springer-Verlag, New York~-- Heidelberg, 1978.

\bibitem{Bog_85}
Bogoyavlensky O.I.,
Methods in the qualitative theory of dynamical systems in astrophysics and gas dynamics, {\it Springer Series in Soviet Mathematics}, Springer-Verlag, Berlin, 1985.

\bibitem{Chan_69}
Chandrasekhar S.,
Ellipsoidal f\/igures of equilibrium, Yale University Press, New Haven~-- London, 1969.

\bibitem{Gold_50}
Goldstein H.,
Classical mechanics, Addison-Wesley Press, Inc., Cambridge, Mass., 1951.

\bibitem{GrNa_91}
Green A.E., Naghdi P.M.,
A thermomechanical theory of a Cosserat point with application to composite materials,
\href{http://dx.doi.org/10.1093/qjmam/44.3.335}{{\it  Quart. J. Mech. Appl. Math.}} {\bf 44} (1991), 335--355.

\bibitem{KovRoz_09}
Kovalchuk V., Ro\.zko E.E.,
Classical models of af\/f\/inely-rigid bodies with ``thickness'' in degenerate dimension,
{\it J. Geom. Symmetry Phys.} {\bf 14} (2009), 51--65,
\href{http://arxiv.org/abs/0902.3573}{arXiv:0902.3573}.

\bibitem{LeSi_90}
Lewis D., Simo J.C.,
Nonlinear stability of rotating pseudo-rigid bodies,
\href{http://dx.doi.org/10.1098/rspa.1990.0014}{{\it Proc. Roy. Soc. London Ser.~A}} {\bf 427} (1990), 281--319.

\bibitem{Love_96}
Love A.E.H.,
A treatise on the mathematical theory of elasticity, Dover, New York, 1996.

\bibitem{NoRe_98}
Nordenholz T.R., O'Reilly O.M.,
On steady motions of isotropic, elastic Cosserat points,
\href{http://dx.doi.org/10.1093/imamat/60.1.55}{{\it IMA J. Appl. Math.}} {\bf 60} (1998), 55--72.

\bibitem{NoRe_01}
Nordenholz T.R., O'Reilly O.M.,
A class of motions of elastic, symmetric Cosserat points: existence, bifurcation, and stability,
\href{http://dx.doi.org/10.1016/S0020-7462(00)00021-4}{{\it Internat. J. Non-Linear Mech.}} {\bf 36} (2001), 353--374.

\bibitem{Roz_05}
Ro\.{z}ko E.E.,
Dynamics of af\/f\/inely-rigid bodies with degenerate dimension,
\href{http://dx.doi.org/10.1016/S0034-4877(05)80087-4}{{\it Rep. Math. Phys.}} {\bf 56} (2005), 311--332.

\bibitem{Roz_PhD}
Ro\.{z}ko E.E.,
Dynamical systems on homogeneous spaces and their applications to continuum mechanics, PhD Thesis, 2006 (in Polish).

\bibitem{Ru_85}
Rubin M.B.,
On the theory of a Cosserat point and its application to the numerical solution of continuum problems,
\href{http://dx.doi.org/10.1115/1.3169055}{{\it J. Appl. Mech.}} {\bf 52} (1985), 368--372.

\bibitem{JJS_82}
S{\l}awianowski J.J.,
The mechanics of the homogeneously-deformable body. Dynamical models with high symmetries,
\href{http://dx.doi.org/10.1002/zamm.19820620604}{{\it Z. Angew. Math. Mech.}} {\bf 62} (1982), 229--240.

\bibitem{JJS_book}
S{\l}awianowski J.J.,
Analytical mechanics of deformable bodies, PWN -- Polish Scientif\/ic Publishers, Warszawa~-- Pozna\'n, 1982 (in Polish).

\bibitem{JJS-VK_03}
S{\l}awianowski J.J., Kovalchuk V.,
Invariant geodetic problems on the af\/f\/ine group and related Hamiltonian systems,
\href{http://dx.doi.org/10.1016/S0034-4877(03)80029-0}{{\it Rep. Math. Phys.}} \textbf{51} (2003), 371--379.

\bibitem{all_04}
S{\l}awianowski J.J., Kovalchuk V., S{\l}awianowska A., Go{\l}ubowska B., Mar\-tens A., Ro\.zko~E.E., Zawistows\-ki~Z.J.,
Af\/f\/ine symmetry in mechanics of collective and internal modes. I.~Classical models,
\href{http://dx.doi.org/10.1016/S0034-4877(04)80026-0}{{\it Rep. Math. Phys.}} {\bf 54} (2004), 373--427,
\href{http://arxiv.org/abs/0802.3027}{arXiv:0802.3027}.

\bibitem{all_05}
S{\l}awianowski J.J., Kovalchuk V., S{\l}awianowska A., Go{\l}ubowska B., Mar\-tens A., Ro\.zko E.E., Zawistows\-ki~Z.J.,
Af\/f\/ine symmetry in mechanics of collective and internal modes. II.~Quantum models,
\href{http://dx.doi.org/10.1016/S0034-4877(05)80002-3}{{\it Rep. Math. Phys.}} {\bf 55} (2005), 1--46,
\href{http://arxiv.org/abs/0802.3028}{arXiv:0802.3028}.

\end{thebibliography}
\end{document}